\documentclass[11pt]{revtex4}   %updated 4 nov 2011%
\usepackage{amssymb}
\usepackage{latexsym}
\usepackage{epsfig}

\begin{document}

\title{Power-law entropy-corrected HDE and NADE in Brans-Dicke cosmology}

\author{A. Sheykhi$^{a,b}$\footnote{sheykhi@mail.uk.ac.ir}, K. Karami$^{c,b}$\footnote{KKarami@uok.ac.ir},
M. Jamil$^{d}$\footnote{mjamil@camp.nust.edu.pk}, E. Kazemi$^{a}$
and M. Haddad$^{a}$}
\address{$^a$Department of Physics, Shahid Bahonar University, P.O. Box 76175, Kerman, Iran\\
         $^b$Research Institute for Astronomy and Astrophysics of Maragha (RIAAM), P. O. Box 55134-441, Maragha,
         Iran\\
         $^c$Department of Physics, University of Kurdistan, Pasdaran St., Sanandaj, Iran\\
         $^d$Center for Advanced
Mathematics and Physics (CAMP), National University of Sciences and
Technology (NUST), Islamabad, Pakistan}

\begin{abstract}
\vspace*{1.5cm} \centerline{\bf Abstract} \vspace*{1cm}
Considering the power-law corrections to the black hole entropy,
which appear in dealing with the entanglement of quantum fields
inside and outside the horizon,  the holographic energy density is
modified accordingly. In this paper we study the power-law
entropy-corrected holographic dark energy in the framework of
Brans-Dicke theory. We investigate the cosmological implications
of this model in detail. We also perform the study for the new
agegraphic dark energy model and calculate some relevant
cosmological parameters and their evolution. {As a result we find
that this model can provide the present cosmic acceleration and
even the equation of state parameter of this model can cross the
phantom line $w_D=-1$ provided the model parameters are chosen
suitably}.

\end{abstract}

 \maketitle
\textit{keywords:} Brans-Dicke; power-law; dark energy.
%---------------------------------------------------------------------------------------------
\section{Introduction\label{Int}}
Recent cosmological and astrophysical data gathered from the
observations of SNe Ia {\cite{c1}}, WMAP {\cite{c2}}, SDSS
{\cite{c3}} and X-ray {\cite{c4}} convincingly suggest that the
observable universe experiences an accelerated expansion phase.
Although the simplest and elegant way to explain this behavior is
the inclusion of Einstein's cosmological constant \cite{c7}, however
 the two deep theoretical problems (namely the ``fine-tuning'' and the
``coincidence'' one) led to the dark energy paradigm. The dynamical
nature of dark energy, at least in an effective level, can arise
from various scalar fields, such as a canonical scalar field
(quintessence) \cite{quint}, a phantom field \cite{phant}, that is a
scalar field with a negative sign of the kinetic term, or the
combination of quintessence and phantom in a unified model named
quintom \cite{quintom}.

One of the dynamic candidates for dark energy is the so-called
``Holographic Dark Energy" (HDE) proposal which is constructed in
the light of the holographic principle. Its origin is the black hole
thermodynamics \cite{BH22} and the connection (known from AdS/CFT
correspondence) of the UV cut-off of quantum field theory, which
gives rise to the vacuum energy, with the largest distance of the
theory \cite{Cohen:1998zx,wang1,pav3}. Thus, determining an
appropriate quantity $L$ to serve as an IR cut-off, imposing the
constraint that the total vacuum energy in the corresponding maximum
volume must not be greater than the mass of a black hole of the same
size, and saturating the inequality, one identifies the acquired
vacuum energy as HDE i.e. $\rho_D=3c^2M_P^2L^{-2}$. Here $M_P$ is
the reduced Planck Mass $M_P^{-2}=8\pi G$. In this model, the
coincidence problem can be resolved by considering a fundamental
assumption that matter and HDE do not conserve separately
\cite{pav1}. It was shown \cite{pav1} that, if there is any
interaction between the two dark components of the universe the
identification of $L$ with Hubble radius, $L=H^{-1}$, necessarily
implies a constant ratio of the energy densities of the two
components regardless of the details of the interaction (see also
\cite{ahmad}). The HDE model has also been tested and constrained by
various astronomical observations where suitable limits have been
obtained on the holographic parameter $c$ and its equation of state
parameter \cite{Xin,Feng}.

It is worthy to note that the entropy-area relationship $S(A)$
yields new results in gravitational physics: for instance, the
entropy-area relation yields the Friedmann equation and the
definition of the HDE. In the later case, the classical relation
$S\sim A \sim L^2$ of black holes yields the dark energy density
$\rho_D=3c^2M_p^2L^{-2}$ \cite{Cohen:1998zx}. The relation $S(A)$
has two interesting modifications (corrections) namely the
logarithmic correction \cite{Wei} and power-law correction
\cite{Sau,pavon1}. In this paper, we are interested in the later
case, that is the modification of HDE due to the power-law
correction to entropy. These corrections arise in dealing with the
entanglement of quantum fields in and out the horizon \cite{Sau}.
The power-law corrected entropy takes the form \cite{pavon1}
\begin{equation}\label{S}
S=\frac{A }{4G}\left(1-K_{\alpha}A^{1-\alpha/2}\right),
\end{equation}
where $\alpha$ is a dimensionless constant whose value is currently
under debate, and
\begin{equation}
\label{K} K_{\alpha}=\frac{\alpha
(4\pi)^{\alpha/2-1}}{(4-\alpha)r_c^{2-\alpha}},
\end{equation}
where $r_c$ is the crossover scale. The second term in (\ref{S}) is
the power-law correction to the entropy-area law. This correction
arises when the wave-function of the field is chosen to be a
superposition of ground state and exited state \cite{Sau}. The
ground state obeys the usual Bekenstein-Hawking entropy-area law.
The corrections to entropy arise only from the excited state, and
more excitations produce more deviation from the entropy-area law
\cite{sau1} (also see \cite{sau2} for a review on power-law entropy
corrections).

Motivated by the power-law entropy corrected relation (\ref{S}), a
new version of HDE called ``power-law entropy-corrected holographic
dark energy" (PLECHDE) was recently proposed \cite{sheyjam}
\begin{equation}\label{rhoD}
\rho _{D }=3c^2M_{P}^{2}L^{-2}-\beta M_{P}^{2}L^{-\alpha}.
\end{equation}
In the special case  $\beta=0$, the above equation yields the
well-known HDE density. When $\alpha=2$ the two terms can be
combined and one recovers again the ordinary HDE density. From
thermodynamical point of view, it was shown \cite{pavon1} that the
generalized second law of thermodynamics for the universe with the
power-law corrected entropy (\ref{S}) is satisfied provided
$\alpha>2$. For $\alpha>2$ the corrected term can be comparable to
the first term only when $L$ is very small \cite{sheyjam}. {Hence,
at the very early stage when the Universe undergoes an inflation
phase, the correction term in the PLECHDE density (\ref{rhoD})
becomes important. When the Universe becomes large, PLECHDE
reduces to the ordinary HDE. Note that after the end of the
inflationary phase, the Universe subsequently enters in the
radiation and then matter dominated eras. In these two epochs,
since the Universe is much larger, the power-law entropy-corrected
term to PLECHDE, namely the second term in Eq. (\ref{rhoD}), can
be safely ignored. Therefore, the PLECHDE can be considered as a
model of entropic cosmology which unifies the early-time inflation
and late-time cosmic acceleration of the Universe.}

Another interesting attempt for probing the nature of dark energy is
the so-called ``agegraphic dark energy'' (ADE). This model was
proposed by Cai \cite{Cai1} to explain the acceleration of the
universe expansion within the framework of quantum gravity. The ADE
model assumes that the observed dark energy comes from the spacetime
and matter field fluctuations in the universe. Since the original
ADE model suffers from the difficulty to describe the
matter-dominated epoch, a new version of ADE was proposed by Wei and
Cai \cite{Wei2}, while the time scale was chosen to be the conformal
time $\eta$ instead  of the age of the universe yielding the
new-agegraphic dark energy (NADE) paradigm. The energy density of
the NADE is given by \cite{Wei2}
\begin{equation}
\rho_{D} = \frac{3n^2{M_P^2}}{\eta^2},\label{ecnade}
\end{equation}
where the conformal time $\eta$ is given by
\begin{equation}
\eta=\int_0^{a}\frac{{d}a}{Ha^2}.\label{eta}
\end{equation}
The agegraphic models of dark energy have been also investigated in
ample details (see e.g \cite{sheage} and references therein). When
the power-law corrections are applied to NADE, the definition
modifies to the form \cite{sheyjam}
\begin{equation}\label{PLECNADE}
\rho_D=3n^2M_{P}^{2}\eta^{-2}-\beta M_P^{2}\eta^{-\alpha}.
\end{equation}
On the other hand, HDE/NADE are dynamical dark energy models, thus
it is more natural to study them in a dynamical framework such as
Brans-Dicke (BD) theory instead of general relativity. Besides, the
BD scalar field speeds up the expansion rate of a dust matter
dominated era (reduces deceleration), while slows down the expansion
rate of cosmological constant era (reduces acceleration). The
studies on the HDE and ADE in the framework of BD theory have been
carried out in  \cite{Pavon2,shey1} and \cite{shey2,karam},
respectively.

For all mentioned above, it is meaningful to investigate the
power-law entropy-corrected HDE/NADE  in the framework of BD theory.
These studies allows us to show the phantom crossing for the EoS
parameter at the present time. We will also study the deceleration
parameter and evolutionary form of dark energy density for both
interacting and noninteracting cases in the BD framework.

This paper is structured as follows. In the next section we study
PLECHDE in BD cosmology. We also calculate the cosmological
parameters of the model. In section III we discuss PLECNADE for both
interacting and noninteracting cases. The last section is devoted to
conclusions.

%---------------------------------------------------------------------------------------------
\section{PLECHDE in Brans-Dicke theory\label{PLECHDE}}

The canonical form of the BD action is given by \cite{Arik}
\begin{equation}
S=\int{
d^{4}x\sqrt{g}\left(-\frac{1}{8\omega}\phi ^2
{R}+\frac{1}{2}g^{\mu \nu}\partial_{\mu}\phi \partial_{\nu}\phi
+L_M \right)},\label{act1}
\end{equation}
where $R$ and $\phi$ are the Ricci scalar and the BD scalar field,
respectively. Taking the variation of action (\ref{act1}) with
respect to the Friedmann-Robertson-Walker (FRW) metric
\begin{eqnarray}
 ds^2=dt^2-a^2(t)\left(\frac{dr^2}{1-kr^2}+r^2d\Omega^2\right),\label{metric}
\end{eqnarray}
yields the Friedmann equations in the framework of BD theory as
\begin{eqnarray}
 &&\frac{3}{4\omega}\phi^2\left(H^2+\frac{k}{a^2}\right)-\frac{1}{2}\dot{\phi} ^2+\frac{3}{2\omega}H
 \dot{\phi}\phi=\rho_D+\rho_M,\label{FE1}\\
 &&\frac{-1}{4\omega}\phi^2\left(2\frac{{\ddot{a}}}{a}+H^2+\frac{k}{a^2}\right)-\frac{1}{\omega}H \dot{\phi}\phi -\frac{1}{2\omega}
 \ddot{\phi}\phi-\frac{1}{2}\left(1+\frac{1}{\omega}\right)\dot{\phi}^2=p_D,\label{FE2}\\
 &&\ddot{\phi}+3H
 \dot{\phi}-\frac{3}{2\omega}\left(\frac{{\ddot{a}}}{a}+H^2+\frac{k}{a^2}\right)\phi=0.
 \label{FE3}
\end{eqnarray}
Here, $\rho_D$ and $p_D$ are the energy density and pressure of dark
energy. Also $\rho_M$ is the energy density of pressureless dark
matter ($p_M=0$). Following \cite{shey1}, we assume $\phi\propto
a^\varepsilon$, then one can get
\begin{eqnarray}\label{dotphi H}
&&\dot{\phi}=\varepsilon H\phi, \\
&&\ddot{\phi}=\varepsilon^2H^2\phi+\varepsilon\phi\dot{H}.\label{ddotphi
H}
\end{eqnarray}
In the framework of BD cosmology, we assume the energy density of
the PLECHDE has the following form
\begin{equation}
\rho _{D }=\frac{3c^2\phi^{2}}{4\omega L^2}-\frac{\beta
\phi^{2}}{4\omega L^{\alpha}},\label{rhoPLHDE}
\end{equation}
where $\phi^2={\omega}/{(2\pi G_{\mathrm{eff}})}$ and
$G_{\mathrm{eff}}$ is the effective gravitational constant. In the
limiting case $G_{\mathrm{eff}}\rightarrow G$, we have
$\phi^2=4\omega M_P^2$ and expression (\ref{rhoPLHDE}) restores the
PLECHDE density in Einstein gravity (\ref{rhoD}). Equation
(\ref{rhoPLHDE}) can be rewritten as
\begin{equation}\label{rhoD in BD H}
\rho _{D }=\frac{3c^2\phi^{2}}{4\omega L^2}\gamma_c,
\end{equation}
where
\begin{equation}
\gamma_c=1-\frac{\beta}{3c^2L^{\alpha-2}}.\label{gammac}
\end{equation}
The IR cut-off $L$ is given by \cite{Huang}
\begin{equation}
L=a(t)\frac{\sin y}{\sqrt{k}},\label{L}
\end{equation}
where $y=\sqrt{k}R_{\rm h}/a$ and
\begin{equation}
R_{\rm h}=a(t)\int_t^\infty \frac{{\rm d} t}{a(t)}=a(t)\int_0^{r}
\frac{{\rm d}r}{\sqrt{1-kr^2}}.\label{y}
\end{equation}
Here $R_{\rm h}$ is the radial size of the event horizon measured in
the $r$ direction and $L$ is the radius of the event horizon
measured on the sphere of the horizon \cite{Huang}.

The critical energy density, $\rho_{\rm cr}$, and the energy density
of curvature, $\rho_k$, are defined as
\begin{eqnarray}\label{rhocr}
\rho_{\mathrm{cr}}=\frac{3\phi^2 H^2}{4\omega},\hspace{0.8cm}
\rho_k=\frac{3k\phi^2}{4\omega a^2}.
\end{eqnarray}
The dimensionless density parameters can also be defined as usual
\begin{eqnarray}
\Omega_M&=&\frac{\rho_M}{\rho_{\mathrm{cr}}}=\frac{4\omega\rho_M}{3\phi^2
H^2}, \label{Omegam H} \\
\Omega_k&=&\frac{\rho_k}{\rho_{\mathrm{cr}}}=\frac{k}{H^2 a^2},\label{Omegak H} \\
\Omega_D&=&\frac{\rho_D}{\rho_{\mathrm{cr}}}=\frac{c^2\gamma_c}{L^2
H^2}.
\label{OmegaD H}
\end{eqnarray}
Using Eqs. (\ref{dotphi H}), (\ref{Omegam H}), (\ref{Omegak H}) and
(\ref{OmegaD H}), one can rewrite the first Friedmann equation
(\ref{FE1}) as
\begin{equation}
1+\Omega_k+2\varepsilon\left(1-\frac{\varepsilon\omega}{3}\right)=\Omega_D+\Omega_M.\label{FE1v2}
\end{equation}
From Eq. (\ref{OmegaD H}) we have
\begin{eqnarray}
HL=\sqrt{\frac{c^2\gamma_c}{\Omega_D}}. \label{HL}
\end{eqnarray}
Taking the derivative of Eq. (\ref{L}) with respect to the cosmic
time $t$ and using (\ref{HL}) yields
\begin{eqnarray}
\dot{L}=\sqrt{\frac{c^2\gamma_c}{\Omega_D}}-\cos y. \label{Ldot}
\end{eqnarray}
Taking the time derivative of Eq. (\ref{rhoD in BD H}) and using
Eqs. (\ref{dotphi H}) and (\ref{Ldot}) we get
\begin{eqnarray}
\dot{\rho}_D=H\rho_D\left[2\varepsilon+\left(1-\frac{1}{c}\sqrt{\frac{\Omega_D}
{\gamma_c}}\cos
y\right)\left(\frac{\alpha-2}{\gamma_c}-\alpha\right)\right]\label{rhodot
H}.
\end{eqnarray}
%---------------------------------------------------------------------------------------------
\subsection{Noninteracting Case}

For the spatially non-flat FRW universe filled with PLECHDE and dark
matter, the energy conservation laws are as follows
\begin{eqnarray}
&&\dot{\rho}_D+3H\rho_D(1+w_D)=0,\label{consq}\\
&&\dot{\rho}_M+3H\rho_M=0, \label{consm}
\end{eqnarray}
where $w_D=p_D/\rho_D$ is the equation of state (EoS) parameter of
PLECHDE. Substituting Eq. (\ref{rhodot H}) in (\ref{consq}), we
obtain immediately the EoS parameter of PLECHDE in BD gravity
\begin{eqnarray}
w_D=-1-\frac{2\varepsilon}{3}-\frac{1}{3\gamma_c}\left(1-\frac{1}{c}\sqrt{\frac{\Omega_D}
{\gamma_c}}\cos y\right)\Big((1-\gamma_c)\alpha-2\Big)\label{wD H}.
\end{eqnarray}
For $\varepsilon=0$ ($\omega\rightarrow\infty$) the BD scalar
field becomes trivial, i.e. $\phi^2=\omega/2\pi G=4\omega M_P^2$,
and Eq. (\ref{wD H}) restores the EoS parameter of PLECHDE in
Einstein gravity \cite{sheyjam}
\begin{eqnarray}
w_D=-1-\frac{1}{3\gamma_c}\left(1-\frac{1}{c}\sqrt{\frac{\Omega_D}
{\gamma_c}}\cos y\right)\Big((1-\gamma_c)\alpha-2\Big)\label{wDH}.
\end{eqnarray}
On the other hand, in the absence of correction term ($\beta=0=\alpha$),
from Eq. (\ref{gammac}) we have $\gamma_c=1$ and Eq. (\ref{wD H})
reduces to the EoS parameter of HDE in BD gravity \cite{shey1}
\begin{equation}
w_D=-\frac{1}{3}-\frac{2\varepsilon}{3}-\frac{2\sqrt{\Omega_D}}{3c}\cos
y.
\end{equation}
For the deceleration parameter
\begin{eqnarray}
q=-\frac{\ddot{a}}{aH^2}=-1-\frac{\dot{H}}{H^2},\label{qdec}
\end{eqnarray}
dividing  Eq. (\ref{FE2}) by $H^2$, and using Eqs. (\ref{dotphi H}),
(\ref{ddotphi H}), (\ref{Omegak H}), (\ref{OmegaD H}) and
(\ref{qdec}), we obtain
\begin{eqnarray}
q=\frac{1}{2\varepsilon+2}\left[(2\varepsilon+1)^2+2\varepsilon(\varepsilon\omega-1)+\Omega_k+3\Omega_D
w_D\right]\label{q1 H}.
\end{eqnarray}
Note that combining the deceleration parameter with the Hubble, the
EoS and the dimensionless density parameters, form a set of useful
parameters for the description of the astrophysical observations.

Replacing Eq. (\ref{wD H}) in (\ref{q1 H}) yields
\begin{eqnarray}
q&=&\frac{1}{2\varepsilon+2}\left[(2\varepsilon+1)^2+2\varepsilon(\varepsilon\omega-1)+\Omega_k
-(2\varepsilon+3)\Omega_D-\frac{\Omega_D}{\gamma_c}\right.
\nonumber\
\\
&& \left.\times\left(1-\frac{1}{c}\sqrt{\frac{\Omega_D}
{\gamma_c}}\cos
y\right)\Big((1-\gamma_c)\alpha-2\Big)\right]\label{q2 H}.
\end{eqnarray}
Again for $\gamma_c=1$ ($\alpha=0$) the above equation reduces to
\cite{shey1}
\begin{eqnarray}
q&=&\frac{1}{2\varepsilon+2}\left[(2\varepsilon+1)^2+2\varepsilon(\varepsilon\omega-1)
+\Omega_k-(2\varepsilon+1)\Omega_D-\frac{2}{c}\Omega_D^{3/2} \cos
y\right]\label{q3 H}.
\end{eqnarray}
%---------------------------------------------------------------------------------------------
\subsection{Interacting Case}
Here, we extend our investigation to the case in which there is an
interaction between PLECHDE and dark matter. The recent
observational evidence provided by the galaxy cluster Abell A586
supports the interaction between dark energy and dark matter
\cite{Bertolami8}. In the presence of interaction, $\rho_D$ and
$\rho_M$ do not conserve separately and the energy conservation
equations become
\begin{eqnarray}
&&\dot{\rho}_D+3H\rho_D(1+w_D)=-Q,\label{consq2H}\\
&&\dot{\rho}_{M}+3H\rho_{M}=Q, \label{consm2H}
\end{eqnarray}
where Q stands for the interaction term. Following \cite{pav1}, we
shall assume
\begin{equation}
Q=3b^2H(\rho_M+\rho_D),\label{Q}
\end{equation}
with the coupling constant $b^2$. Substituting Eqs. (\ref{rhodot H})
and (\ref{Q}) in Eq. (\ref{consq2H}) and using (\ref{FE1v2}) gives
\begin{eqnarray}
w_D=-1-\frac{2\varepsilon}{3}-\frac{1}{3\gamma_c}\left(1-\frac{1}{c}\sqrt{\frac{\Omega_D}
{\gamma_c}}\cos y\right)\Big((1-\gamma_c)\alpha-2\Big)
\nonumber\\
-\frac{b^2}{\Omega_D}\left[1+\Omega_k+2\varepsilon\left(1-\frac{\varepsilon\omega}{3}\right)\right]
\label{wDINTH}.
\end{eqnarray}
For $\gamma_c=1$ ($\alpha=0$), the above equation reduces to the EoS
parameter of interacting HDE in BD cosmoligy \cite{shey1}
\begin{eqnarray}
w_D=-\frac{1}{3}-\frac{2\varepsilon}{3}-\frac{2\sqrt{\Omega_D}}{3c}\cos
y-\frac{b^2}{\Omega_D}\left[1+\Omega_k+2\varepsilon\left(1-\frac{\varepsilon
\omega}{3}\right)\right]\label{wDInt1 H}.
\end{eqnarray}
In the presence of interaction, the deceleration parameter for
PLECNADE model can be obtained by replacing Eq. (\ref{wDINTH}) in
(\ref{q1 H}) as
\begin{eqnarray}
q&=&\frac{1}{2\varepsilon+2}\left[(2\varepsilon+1)^2+2\varepsilon(\varepsilon\omega-1)
+\Omega_k-(2\varepsilon+3)\Omega_D-\frac{\Omega_D}{\gamma_c}\right.
\nonumber\
\\
&& \left.\times\left(1-\frac{1}{c}\sqrt{\frac{\Omega_D}
{\gamma_c}}\cos y\right)\Big((1-\gamma_c)\alpha-2\Big)
-3b^2\left[1+\Omega_k+2\varepsilon\left(1-\frac{\varepsilon\omega}{3}\right)\right]
\right]\label{q2 H}.
\end{eqnarray}
Again for $\gamma_c=1$ ($\alpha=0$) we have \cite{shey1}
\begin{eqnarray}
q&=&\frac{1}{2\varepsilon+2}\left\{(2\varepsilon+1)^2+2\varepsilon(\varepsilon\omega-1)
+\Omega_k-(2\varepsilon+1)\Omega_D-\frac{2}{c}{\Omega^{3/2}_D}\cos
y\right. \nonumber\
\\
&&
\left.-3b^2\left[1+\Omega_k+2\varepsilon\left(1-\frac{\varepsilon\omega}{3}\right)\right]\right\}\label{q2Int
H}.
\end{eqnarray}
Taking time derivative of Eq. (\ref{HL}) and using
${\dot{\Omega}_D}=H{\Omega'_D}$, one can get the equation of motion
for $\Omega_D$ as
\begin{eqnarray}\label{OmegaDp H}
\Omega'_D=\Omega_D\left[2+2q+\left(\frac{\alpha-2}
{\gamma_c}-\alpha\right)\left(1-\frac{1}{c}\sqrt{\frac{\Omega_D}{\gamma_c}}\cos
y\right)\right],
\end{eqnarray}
where the prime denotes the derivative with respect to $x=\ln a$ and
$q$ is given by Eq. (\ref{q2 H}). In the absence of correction we
have $\gamma_c=1$ ($\beta=0=\alpha$), thus Eq. (\ref{OmegaDp H})
restores \cite{shey1}
\begin{eqnarray}
{\Omega'_D}=2\Omega_D\left(q+\frac{\sqrt{\Omega_D}}{c}\cos y
\right).
\end{eqnarray}
%---------------------------------------------------------------------------------------------
\section{PLECNADE in Brans-Dicke theory\label{PLECNADE}}

The PLECNADE density in BD gravity is given by
\begin{equation}\label{OmegaD}
\rho _{D }=\frac{3n^2\phi^{2}}{4\omega \eta^2}-\frac{\beta
\phi^{2}}{4\omega \eta^{\alpha}}.
\end{equation}
We rewrite Eq. (\ref{OmegaD}) as
\begin{equation}\label{rhoD in BD}
\rho _{D }=\frac{3n^2\phi^{2}}{4\omega \eta^2}\gamma_n,
\end{equation}
with
\begin{equation}\label{gamman}
\gamma_n=1-\frac{\beta}{3n^2\eta^{\alpha-2}}.
\end{equation}
From definition
$\rho_D=\Omega_D\rho_{\mathrm{cr}}=3\phi^2H^2\Omega_D/(4\omega)$ and
using Eq. (\ref{rhoD in BD}), we get
\begin{eqnarray}
H\eta=\sqrt{\frac{n^2\gamma_n}{\Omega_D}}.\label{Heta}
\end{eqnarray}
Taking time derivative of Eq. (\ref{rhoD in BD}), using (\ref{dotphi
H}) and $\dot{\eta}=1/a$ we obtain
\begin{eqnarray}
\dot{\rho}_D=H\rho_D\left[2\varepsilon+\frac{1}{an}\sqrt{\frac{\Omega_D}
{\gamma_n}}\left(\frac{\alpha-2}{\gamma_n}-\alpha\right)\right]\label{rhodot}.
\end{eqnarray}

\begin{figure}[h]
\centering
\includegraphics{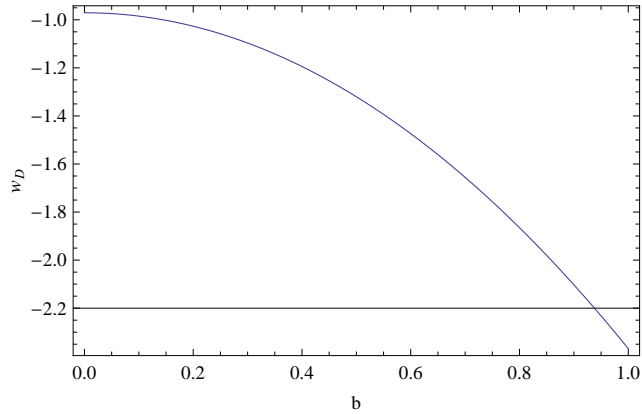}
\caption{The equation of state parameter of PLECNADE is plotted
against interacting coupling parameter $b$. The trial values of
model parameters are $\omega=10^{4}$, $\omega\epsilon\approx1$,
$\alpha=2.8$, $\Omega_k=0.02$, $n=2.716$, $a=1$, $\gamma_n=100$,
$\Omega_D=0.73$.} \label{fig2}
\end{figure}

%---------------------------------------------------------------------------------------------
\subsection{Noninteracting Case}

For noninteracting PLECNADE in BD theory, the EoS parameter can be
obtained by replacing Eq. (\ref{rhodot}) in (\ref{consq}). The
result is
\begin{eqnarray}
w_D=-1-\frac{2\varepsilon}{3}-\frac{1}{3an}\sqrt{\frac{\Omega_D}
{\gamma_n}}\left(\frac{\alpha-2}{\gamma_n}-\alpha\right)\label{wDn}.
\end{eqnarray}
In the special case $\gamma_n=1$ $(\alpha=0)$, Eq. (\ref{wDn})
restores the EoS parameter of NADE in BD gravity
\cite{shey2}
\begin{equation}
w_D=-1-\frac{2\varepsilon}{3}+\frac{2\sqrt{\Omega_D}}{3na}.
\end{equation}
The deceleration parameter $q$ for the noninteracting PLECNADE model
is still obtained according to Eq. (\ref{q1 H}), where $w_D$ is now
given by Eq. (\ref{wDn}). Substituting Eq. (\ref{wDn}) in (\ref{q1
H}) gives
\begin{eqnarray}
q&=&\frac{1}{2\varepsilon+2}\left[(2\varepsilon+1)^2+2\varepsilon(\varepsilon\omega-1)
+\Omega_k-(2\varepsilon+3)\Omega_D-\frac{{\Omega_D}^{3/2}}{an\sqrt{\gamma_n}}\right.
\nonumber\
\\
&& \left. \times \left(\frac{\alpha-2}{\gamma_n}-\alpha\right)
\right]\label{q2}.
\end{eqnarray}

\begin{figure}[h]
\centering
\includegraphics{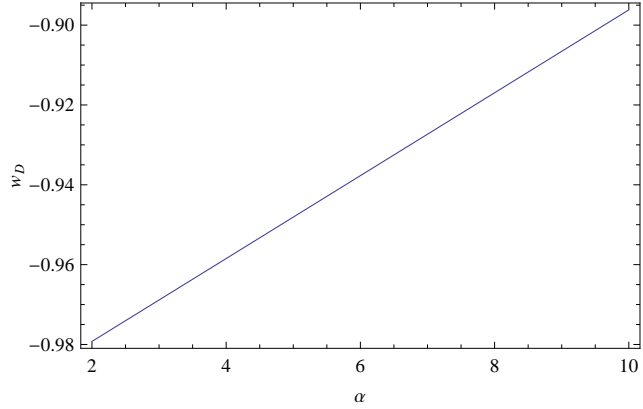}
\caption{The equation of state parameter of PLECNADE (Eq.53) is
plotted against power-law parameter $\alpha$. The trial values of
model parameters are $\omega=10^{4}$, $\omega\epsilon\approx1$,
$\Omega_k=0.02$, $n=2.716$, $a=1$, $b=0.01$, $\gamma_n=100$,
$\Omega_D=0.73$.} \label{fig3}
\end{figure}

%---------------------------------------------------------------------------------------------
\subsection{Interacting Case}

Here, the EoS parameter of interacting PLECNADE in BD
gravity is obtained by replacing Eqs. (\ref{Q}) and (\ref{rhodot})
into Eq. (\ref{consq2H}) and using (\ref{FE1v2}). The result yields
\begin{eqnarray}
w_D=-1-\frac{2\varepsilon}{3}-\frac{1}{3an}\sqrt{\frac{\Omega_D}
{\gamma_n}}\left(\frac{\alpha-2}{\gamma_n}-\alpha\right)
-\frac{b^2}{\Omega_D}\left[1+\Omega_k+2\varepsilon\left(1-\frac{\varepsilon\omega}{3}\right)\right]\label{wDINT}.
\end{eqnarray}
For $\gamma_n=1$ ($\beta=0=\alpha$), the above equation reduces to
\cite{shey2}
\begin{equation}
w_D=-1-\frac{2\varepsilon}{3}+\frac{2\sqrt{\Omega_D}}{3na}
-\frac{b^2}{\Omega_D}\left[1+\Omega_k+2\varepsilon\left(1-\frac{\varepsilon\omega}{3}\right)\right].
\end{equation}
The deceleration parameter is obtained by replacing Eq.
(\ref{wDINT}) in (\ref{q1 H})
\begin{eqnarray}
q&=&\frac{1}{2\varepsilon+2}\left\{(2\varepsilon+1)^2+2\varepsilon(\varepsilon\omega-1)
+\Omega_k-(2\varepsilon+3)\Omega_D-\frac{{\Omega_D}^{3/2}}{an\sqrt{\gamma_n}}\right.
\nonumber\
\\
&& \left. \times\left(\frac{\alpha-2}{\gamma_n}-\alpha\right)
-3b^2\left[1+\Omega_k+2\varepsilon\left(1-\frac{\varepsilon\omega}{3}\right)\right]
\right\}\label{q2}.
\end{eqnarray}
Taking time derivative of Eq. (\ref{Heta}), using
${\dot{\Omega}_D}=H{\Omega'_D}$ and $\dot{\eta}=1/a$, the equation
of motion for $\Omega_D$ can be obtained as
\begin{eqnarray}\label{OmegaDp}
\Omega'_D=\Omega_D\left[2+2q+\frac{1}{an}\sqrt{\frac{\Omega_D}
{\gamma_n}}\left(\frac{\alpha-2} {\gamma_n}-\alpha\right)\right],
\end{eqnarray}
which in the absence of correction term $\gamma_n=1$ ($\alpha=0$)
reduces to the result obtained for NADE in BD gravity
\cite{shey2}
\begin{eqnarray}\label{OmegaDp}
\Omega'_D=2\Omega_D\left[1+q-\frac{1}{an}\sqrt{\Omega_D }\right].
\end{eqnarray}

In figures 1 and 2, we have plotted the equation of state
parameter of  PLECNADE against various model parameters. From
these figures one can see explicitly that the equation of state
parameter can cross the phantom boundary $w_D=-1$, thus realizing
the phenomenon of cosmic acceleration.

%---------------------------------------------------------------------------------------------
\section{Conclusions}
In this paper, we investigated the models of HDE and NADE with
power-law correction taking a non-flat FRW background in the BD
gravitational theory. The power-law correction is motivated from
the entanglement of quantum fields in and out the horizon. The BD
theory of gravity involves a scalar field which accounts for a
dynamical gravitational constant. We assumed an ansatz by which
the BD scalar field evolves with the expansion of the universe. We
then established a correspondence between the field and the
PLECHDE (and PLECNADE) to study its dynamics. The dynamics are
governed by few dynamical parameters like its equation of state,
deceleration parameter and energy density parameter. For the sake
of generality, we calculated them in the non-flat background with
the interaction of PLECHDE (and PLECNADE) with the matter.
Interestingly  enough we found that the presented model can
accomodate the phantom regim for the equation of state parameter
provided the model parameters are chosen suitably. To clarify this
point we plotted the evolution of $w_D$ against scale factor and
demonstrated explicitly that cosmic acceleration and phantom
crossing can be realized in our model.

%%%%%%%%%%%%%%%%%%%%%%%%%%%%%%%%%%%%%%%%%%%%%%%%%%%%%%%%%%%%%%
\acknowledgments{We are grateful to the referee for valuable
comments and suggestions, which have allowed us to improve this
paper significantly. The works of A. Sheykhi and K. Karami have
been supported financially by Research Institute for Astronomy and
Astrophysics of Maragha (RIAAM) under research project No. 1/2338
Iran. }
%%%%%%%%%%%%%%%%%%%%%%%%%%%%%%%%%%%%%%%%%%%%%%%%%%

%%%%%%%%%%%%%%%%%%%%%%%%%%%%%%%%%%%%%%%%%%%%%%%%%

\end{document}